\documentclass[sigplan,screen]{acmart}
\usepackage{graphicx}
\usepackage{todonotes}
\usepackage{listings}
\usepackage{float}
\input{dafny.sty}
\begin{document}
\newcommand{\acignore}[1]{{}}
\title{Towards the verification of a generic interlocking logic:\\
Dafny meets parameterized model checking}

\author{Alessandro Cimatti}
\affiliation{%
  \institution{Fondazione Bruno Kessler (FBK)}
  \country{}
}
\email{cimatti@fbk.eu}

\author{Alberto Griggio}
\affiliation{%
 \institution{Fondazione Bruno Kessler (FBK)}
  \country{}
}
\email{griggio@fbk.eu}

\author{Gianluca Redondi}
\affiliation{
 \institution{Fondazione Bruno Kessler (FBK)}
  \country{}
}
\email{gredondi@fbk.eu}


\begin{abstract}
Interlocking logics are at the core of critical systems controlling the traffic within stations. In this paper, we consider a generic interlocking logic, which can be instantiated to control a wide class of stations.
We tackle the problem of parameterized verification, i.e. prove that the logic satisfies the required properties for all the relevant stations.
We present a simplified case study, where the interlocking logic is directly encoded in Dafny.
Then, we show how to automate the proof of an important safety requirement, by integrating simple, template-based invariants and more complex invariants obtained from a model checker for parameterized systems.
Based on these positive preliminary results, we outline how we intend to integrate the approach by extending the IDE for the design of the interlocking logic.
\end{abstract}
\maketitle

\section{Introduction}

Interlocking systems are complex, safety-critical systems controlling the operation of the devices in a railway station. The main function is the creation of safe routes for trains from different points in the station. This requires that the devices insisting upon a given route (e.g. semaphores, switches, level crossing) must be properly operated and that mutual exclusion between interfering routes is ensured.
At a high level of abstraction, an interlocking system can be thought of as implementing a very articulated mutual exclusion protocol, that we refer to as interlocking logic.

In this paper, we investigate the use of a formal approach to ensure the reliability and integrity of their operations, as a complement to the standard validation and certification techniques. 
%
%
Our context is an ongoing activity of RFI (the company managing the Italian railways network), aiming to develop an in-house, framework to design interlocking logics and support the development of interlocking systems~\cite{IDE, IDE2}.
One of the goals of this framework is to develop a computer-based, in order to go beyond the current relay-based interlocking technology.
%
Starting from a high-level controlled natural language to describe the interlocking logic, a model-based Integrated Development Environment supports the railways signaling engineers in specifying the interlocking logic in a well-structured and semantically unambiguous way.
Interestingly, the interlocking logic is \emph{generic} in that it describes the procedures without specific reference to a single, given station; rather, it applies to any station in a given class.
The resulting specification is then translated into a SysML model, and from there compiled into executable code.
Then, the user can define a specific station configuration (e.g. the station of the city of Trento), detailing the exact number of components and their interactions.
Upon configuration, the code can be tested in a closed loop integration with a yard simulator modeling the behavior of trains and physical devices.

Formally verifying the interlocking logic is a very important goal, and several attempts have been made in this direction \cite{Fantechi1, FMRailway}. On the one side, we attempted to apply software model checking techniques on the code configured with respect to a specific station. This approach hit a scalability barrier, due to the sheer size of the resulting, instantiated model. Even more importantly, the results of the verification would be applicable to a specific station only.
On the other side, we tried to tackle the problem of verifying the generic interlocking logic without instantiating it with respect to a specific station. We attempt to follow an SMT-based approach aiming at devising quantified invariants. This leads to large and difficult-to-reason-about formulae, due to the lack of a proper structure.

Based on these experiences, we are investigating an alternative approach, that leverages the object-oriented features of Dafny to obtain a high-level and natural framework where we tackle the verification of generic interlocking logics. We present an artificial but representative case study, that shows how we intend to model the problem in Dafny and show how we verify some generic properties. We hint that automation in the verification can be increased by combining template-based generation of simple invariants and the use of a  parameterized model checker to generate more complex, problem specific invariants. We believe that the results are quite promising and that the approach is worth investigating further. We discuss the open challenges that we still need to solve to make this approach scalable.

\section{Developing a Generic Interlocking Logic}

In this section, we overview the approach for the development of a generic interlocking logic~\cite{IDE,IDE2,Norma}.
The development of the interlocking logic is model-based and starts from a domain-specific controlled natural language (CNL), supported by the tool AIDA. 
The interlocking logic is designed through the creation of \emph{sheets}, each defining a logical entity, also referred to as a class.
Examples of entities include shunting routes, i.e. the high-level process devoted to creating a safe path for a train within the station, and lower level entities such as track segments, switches, level crossing, axal counters, and semaphores.
Each sheet is divided into two parts. The first one defines the structure of the class, i.e. variables, parameters, and notably lists of other components that are connected to the class. For example, a shunting route will have lists of the entities it track segments it insists upon, e.g. the track segments that must be locked before the green light is signaled.

The second part of the sheet describes the behavior of the component, which can be thought of as an extended Finite State Machine.
A distinguished state variable, taking values from an enumerative set, is used to define the current location in the FSM.
Each state transition is characterized by the following elements:
\begin{itemize}
\item Source and Destination;
\item Triggers, i.e. events that determine the firing of a transition (e.g. the reception of a command from an external operator, such as "create the route from entrance 3 to platform 12");
\item Guards, i.e. the conditions that must be satisfied to enable a particular transition. Determinism is ensured by explicitly prioritizing guards; 
\item Effects: when a transition is executed, effects are applied that alter the internal state of the component, and possibly the state of components that are connected to it.
\end{itemize}
Guards and effects of a transition may read and or write the value of variables of the objects in the list of the connected components connected to the class. 

The structured natural language used in the sheets has been designed to be comprehensible even to those not trained in formal languages and incorporates grammatical structures drawn from domain-specific jargon. Phrases are structured to ensure traceability to pertinent provisions and regulations. For instance, an engineer might specify a transition guard as follows: "Check that all the track segments of the routes are in a free state."

In AIDA, the sheets written in controlled natural language are associated with a number of syntactic and semantic checks and are translated into a SysML model.
From the SysML model, it is possible to extract graphical views of the FSM for each class.
Furthermore, the SysML can be compiled into executable code (Python and C).

The IDE is complemented by two other tools: TOSCA and Norma.
TOSCA implements a number of functions for the testing of the interlocking logic. It supports the specification of test scenarios, using a domain specific CNL sharing some features with the CNL for the interlocking logic. Generic, abstract scenarios may refer to any specific station and can be automatically instantiated on concrete scenarios. These can then be run on the suitably instantiated software, in closed loop with a simulator of the trains and trackside devices. Automated test case generation oriented to coverage is also supported.

Norma\cite{Norma} is the tool for the formal analysis of legacy interlocking systems based on relay technology. The role of Norma is to extract formulae from the old designs and test that the new specifications will reflect the behavior of the older systems. The definition of a semantic correspondence between the two interlocking technologies is nontrivial and relies on the definition of specific abstractions~\cite{Reverseing} to extract properties to be used in verification.

Within this comprehensive IDE, preliminary experiments in applying formal verification have been attempted, aiming at proving safety properties of the generated C code. The tool leverages symbolic model-checking techniques for software verification, with integration into the Kratos2 model checker~\cite{Kratos2}.
This approach, however, did not yield the expected results.
On the one hand, even if the generation of C code is generic to all 
configurations, its subsequent verification can only take place once 
the configuration of a specific station has been provided.
This is necessary to meet the restrictions of Kratos2, that is unable
to deal with objects of unspecified size.
%
As a result, our current capability allows us to assess 
the safety of individual stations with regard to specific properties.
Furthermore, the verification of the C implementation of the interlocking logic configured for a given station incurs scalability problems, due to the large number of components and their complex connections.
For this goal, we could use the generic generated code as an input for Kratos2, but the invariant generation engine of the model checker is not able to synthesize the correct parameterized invariants.
Hence, in the rest of this paper, we tackle the parameterized verification of the \emph{generic} interlocking logic, without assuming that a specific station configuration is given. 

\section{Dafny encoding}

The features of the Dafny language support a  natural encoding of the interlocking logic described in previous section. Moreover, the verification in Dafny is modular, 
thus possibly reducing the complexity of the system to verify the single components. 
We now illustrate a possible Dafny implementation of a simple example, that has been designed to be representative and comprehensive of all the constructs of the interlocking logic. In an ideal scenario, this Dafny code will be generated automatically from SysML within the IDE, complementing the generation of C and Python code. 
Our example consists of a simple station with  two kinds of components: tracks, representing the track segments, and routes, modeling the different routes for the possible trains. 

The track class has two variables: a \textit{state} variable, and a \textit{direction} variable, that is used to simulate the presence of switches.
Similarly, the routes have a state variable and a Boolean one. In addition, each route is connected to a set of tracks (the ones used by the route), which should be in a particular direction.

Regarding transitions, we consider for simplicity only one simple method of the route class, that mirrors the transition from an inactive state to an active one. The guard corresponds to the condition `all the tracks of the route are in a free state and in the requested direction', and the effect is `assign to all the tracks of the route the state locked'.

In the figures \ref{fig:track-class-dafny} and \ref{fig:route-class-dafny} we can see the two Dafny definitions of the classes. We start by defining variables, as we would find them in the initial segment of an AIDA sheet. Subsequently, the methods of the class model the transitions in the second part of the AIDA sheet.
The source and destination states, guards, and effects of transitions are translated into a series of preconditions and postconditions for each method. 
The body of each method is populated with a representation of the C code generated by AIDA. 
\begin{figure}[h]
\begin{lstlisting}[language=dafny, frame=single]
datatype TrackState = Init | Locked | Free
datatype TrackDirection = Straight | Reverse


class Track
{
    var state : TrackState 
    var direction : TrackDirection

    constructor (d : TrackDirection)
        ensures this.state == TrackState.Init
        ensures this.direction == d
    {
        state := TrackState.Init;
        direction := d;        
    }

    method lock()
        modifies this
        ensures this.state == Locked
        ensures this.direction == old(this.direction)
    {
        this.state := Locked;
    }

    method go_free()
        modifies this
        ensures this.state == Free
        ensures this.direction == old(this.direction)
    {
        this.state := Free;
    }

    method change_dir()
        modifies this
        ensures this.state == old(this.state)
        ensures old(this.direction) != this.direction 
    {
        if (this.direction == Straight)
            {
                this.direction := Reverse;
            }
        else {
            this.direction := Straight;
        }
    }

}
\end{lstlisting}
\caption{Dafny encoding of the track class}
\label{fig:track-class-dafny}
\end{figure}
 
\begin{figure}[h]
\begin{lstlisting}[language=dafny, frame=single]
include "tracks.dfy"

datatype RouteState = Init | Active | Wait | Inactive

class Route 
{      
    var state : RouteState
    var change_dir_request : bool

    const used : seq<Track> 
    const requested_dir : seq<TrackDirection>

    ghost predicate ValidRoute()
        reads this
        {
            |this.used| == |this.requested_dir|
        }

    constructor (u : seq<Track>, d: seq<TrackDirection>)
        requires |u| == |d|
        ensures this.state == RouteState.Init
        ensures this.requested_dir == d
        ensures this.used == u
        ensures change_dir_request == false
        ensures ValidRoute()
    {
        used := u;
        requested_dir := d;
        state := RouteState.Init;
        change_dir_request := false;
    }
        
    method activateroute()
        modifies this, this.used[..]
        requires ValidRoute()
        requires this.state == Wait
        requires forall t :: t in this.used ==> t.state == Free
        requires forall i :: 0 <= i < |this.used| ==> 
            this.used[i].direction == this.requested_dir[i]
        ensures ValidRoute()
        ensures this.state == Active
        ensures this.change_dir_request == 
            old(this.change_dir_request)
        ensures (forall t :: t in this.used 
            ==> t.state == Locked)
        ensures (forall t :: t in this.used 
            ==> t.direction == old(t.direction)) 
        {
            var i := 0;
            while i < |this.used|
                modifies this.used[..]
                invariant 0 <= i <= |this.used|
                invariant forall t :: t in this.used[..i] 
                    ==> t.state == Locked    
                invariant forall t :: t in this.used 
                    ==> t.direction == old(t.direction)
            {
                this.used[i].lock();
                i := i + 1;
            }  

            this.state := Active;
        } 
}



\end{lstlisting}
\caption{Dafny encoding of the route class}
\label{fig:route-class-dafny}  
\end{figure}
We remark that, unlike the description of the AIDA sheet, the formalization of the methods in Dafny introduces a "modifies" clause. Additionally, we must explicitly include equalities among variables that remain unchanged in the postconditions. The initial challenge in automating Dafny code generation lies in accurately representing these statements.

Up to this point, the verification checks performed by Dafny are focused on ensuring that the generated code complies with the specified requirements. The only invariants required for this verification occur in the while loop of the `activate route' method. We suppose that we can generate such invariants automatically, without the need of an external tool, as they reflect the structure of the generated code, which is under our control.

Lastly, we introduce the concluding segment of the Dafny code designed for verifying the safety of the station, in figure \ref{fig:station-dafny}. 
\begin{figure}
\begin{lstlisting}[language=dafny, frame=single]
include "tracks.dfy"
include "routes.dfy"

class Station
{
    const routes : set<Route>
    const tracks : set<Track>

    ghost predicate ValidStation()
     reads this, this.routes
    {
        && this.routes != {}
        && this.tracks != {}
        && (forall i :: i in this.routes ==> 
            (exists j :: j in this.tracks && j in i.used))
        && (forall i, t :: i in this.routes && 
            t in i.used ==> t in this.tracks)
        && forall r :: r in routes ==> r.ValidRoute()
    }
    ghost predicate NotCompatibleRoutes(a : Route, 
        b : Route)
        reads a, b
    {
        (exists i :: (i in a.used) && (i in b.used)) && a != b
    }
        
    ghost predicate SecureStation ()
        reads this, this.routes
    {
      forall i,j  :: (i in this.routes) && 
        (j in this.routes) && (NotCompatibleRoutes(i,j)) 
        ==> !(i.state == Active && j.state == Active)
    }

    ghost predicate SecureInductiveStation() 
        reads this, this.tracks, this.routes
    {
        this.SecureStation() && 
        forall r, t :: (r in this.routes && t in this.tracks
             && t in r.used && t.state == Free) 
                ==> ! (r.state == Active)
    }      

    predicate Precondition(r : Route)
        requires r.ValidRoute()
        reads r, r.used
    {
        && (forall t :: t in r.used 
                        ==> t.state == Free)
        && (forall i :: 0 <= i < |r.used|
                ==> r.used[i].direction ==
                     r.requested_dir[i])
    }

    method StationScheduler()
        modifies this.routes, this.tracks
        requires ValidStation()
        requires SecureInductiveStation()
        ensures ValidStation()
        ensures SecureInductiveStation()
        decreases * 
    {
        while true
            modifies this.routes, this.tracks
            invariant ValidStation()
            invariant SecureInductiveStation()
            decreases *
            {
                var r :| r in this.routes;
                match r.state {
                    case Init => r.deactivate();
                    case Inactive => r.requestroute();
                    case Wait => if Precondition(r)
                            {r.activateroute();}
                    case Active => r.deactivateroute()}
            }
    }

}
\end{lstlisting}
\caption{Dafny encoding of the station}
\label{fig:station-dafny}
\end{figure}
Within this last code segment, we introduce the concept of a station, which is portrayed as a collection of routes and tracks. Subsequently, we define the station's scheduler, a crucial element in the system. This scheduler nondeterministically selects a method that can be executed and proceeds with its execution. 

Additionally, we define the property that we seek to validate, named "Secure Station," which corresponds to the property `two incompatible routes cannot be active together'.

The loop of the scheduler tries to establish that the property is preserved by each method. Notably, this preservation is not true for the initial property itself, "Secure Station." However, the verification succeeds when we provide a stronger inductive invariant that implies the original property - called "Secure Inductive Station".

The goal of this project would be to use a parameterized model checker to find these inductive invariants automatically.

\section{Invariant inference with a Parameterized Model Checker}


Parameterized model checking is a verification technique designed to assess whether a given property holds true for all possible system configurations, making it particularly valuable for systems with variable sizes or structures. This approach allows for the analysis of system behavior, safety, and correctness across a range of instances, providing a formal framework for universal reliability. 
Unfortunately, with the exception of a few cases, parameterized verification is undecidable. Nonetheless, in the literature, there are various approaches that can automatically synthesize invariants for parameterized systems \cite{MCMT, Lambda, cade, Abdulla2018}. In particular, the algorithm presented in \cite{Lambda} is an SMT-based algorithm for the verification of parameterized systems. In that setting, unbounded components are modeled via a theory of a simple type, and state variables are functions from such types to other theories. 

Our goal is to use such a model checker to synthesize automatically the inductive invariants that Dafny needs to conclude the proof. We illustrate the main concepts of SMT-based parameterized model checking continuing the last example.
To formalize the whole system symbolically, we need to define a transition system $S = (X, I(X), T(X, X'))$, where $X$ is a set of variables, and $I, T$ are formulas over some theory. A key insight here is that we do not need to represent all the attributes of the class, but only the one relevant to building the inductive invariant. In this example, we have obtained this simplification manually, by selecting the variables that were more significant. Ideally, such a selection should be done automatically by some kind of slicing abstraction, guided by domain-based intuitions.

For this example, the only variables we need are two \textit{state} variables, one with sort $track \mapsto \{free, locked\}$ and one with sort $route \mapsto \{active, inactive \}$. The initial formula of the system is:
$$\forall t : track.(state[t] =  free) \wedge \forall r : route.(state[r] = inactive). $$
Moreover, we model the dependencies between a route and its associated tracks with a binary predicate $UsedBy$ with sort $ route \times state \mapsto Bool$. The method `activate route' can thus be modeled with the following formula: 
\begin{multline*}
\exists r : route \big( state[r] = inactive  \wedge \\
\forall t1 : track (Usedby(t1, r) \rightarrow state[t1] = free) \wedge \\
{} \wedge state'[r] = active \wedge \forall s : route (s \neq r \rightarrow state'[s] = state[s]) 
\\{}\wedge \forall t : track. ( Usedby(t, r) \rightarrow state'[t] = locked \\
\wedge \neg Usedby(t, r) \rightarrow state'[t] = state[t]) \big) .
\end{multline*}
Then, we define
\begin{multline*}
  NotCompatible(r1, r2) \iff \\
  r1 \neq r2 \wedge \exists t : track. Usedby(t, r1) \wedge Usedby(t, r2),
\end{multline*}
and the candidate property, represented by the formula:
\begin{multline*}
\forall r1 , r2 : route \big( NotCompatible(r1, r2) \rightarrow \\
\neg (state[r1] = active \wedge state[r2] = active)  \big).
\end{multline*}

using the (implementation of the) algorithm of~\cite{Lambda},
we can find (in less than one second) the \textit{lemma}
\begin{multline*}
\forall r : route, t : track ( Usedby(t, r) \wedge state[t] = free) \\
                \rightarrow (state[r] \neq active)
\end{multline*}
which, in conjunction with the original property, is an inductive invariant for $S$. This is the same invariant as the `Secure Inductive Station' in the Dafny code of the station. 
It's important to highlight that the symbolic system $S$ represents a simplified version of the station. In this simplified example, both tracks and routes contain fewer variables compared to their original counterparts.

This straightforward illustration portrays an ideal scenario where Dafny and a parameterized model checker seamlessly collaborate to arrive at a conclusive proof. Without Dafny, the typical approach would require monolithic use of the model checker, but this approach often struggles to scale effectively when dealing with exceptionally large models. 
\section{Ongoing and Future Work}

We tackle the problem of formally verifying an interlocking logic expressed in a domain specific language. The main problem is that the logic is parameterized, in the sense that it is intended to control any station with an arbitrary number of components.
We preliminarily analyzed a highly simplified case study, with two main insights. First, we confirm that it is possible to directly encode the main features of the interlocking logic in Dafny in a very natural way. Second, we investigate verification and the relation to simple invariants (from predefined schemata) and to more complex invariants (resulting from the application of parameterized model checker~\cite{Lambda}).
%

Given the successful preliminary steps, we intend to extend the IDE for the Interlocking logic to support parameterized verification. The first step will be to devise an encoder to automatically generate Dafny code automatically from SysML. We expect this step to be relatively simple, given that the interlocking logic constructs have a direct correspondence to Dafny ones, and back-and-forth traceability can be achieved.
The Dafny code will incorporate both 
the method bodies, mirroring the C and Python code, and the preconditions and postconditions, echoing the engineers' natural language specifications.
Second, we will integrate a way to express the properties to be proved, likely leveraging the language for specifying the abstract scenarios in the TOSCA environment~\cite{IDE}.
Third, we will integrate the generation of invariants required to show that the interlocking logic satisfies the expected properties. On the one side, we will instrument the encoding to automatically generate "simple" invariants via templates, to check the compliance of the generated code to specification. On the other, we will integrate a parameterized model checker (and possibly other invariant generators) to infer invariants for general safety properties of the interlocking logic.

For the latter, we anticipate a more complex path. Parameterized model checkers are not designed for tackling large-scale problems. Hence, our plan is to utilize them on smaller abstractions derived from the code. After the abstraction is built, we are not interested in finding immediately an invariant: even if the parameterized model checker takes hours to synthesize a good invariant, we would be satisfied. However, the most hard challenge will likely be finding the correct abstractions. It could be possible to explore the notion of guiding this abstraction process with insights gained from failed verifications. Extracting such insights from the Dafny verification conditions may be arduous, as the formulae generated by Boogie might be cryptic. As a result, we think that the design of the abstraction-refinement loop should be conducted outside the realm of Dafny.

Furthermore, we do not dismiss the idea of a semi-automated approach, where railway engineers can contribute lemmas or provide guidance in the abstraction process, 
potentially in controlled natural language.

\bibliographystyle{ACM-Reference-Format}
\bibliography{ref}

\end{document}